\documentclass[runningheads]{cl2emult}

\usepackage{makeidx}  
\usepackage{graphicx} 
\usepackage{subeqnar} 
\usepackage{multicol} 
\usepackage{cropmark} 
\usepackage{lnp}      
\makeindex            

\begin{document}
\title*{A local infrared perspective to deeper ISO surveys}
\toctitle{A local infrared perspective to deeper ISO surveys}
%
%
\titlerunning{A local infrared perspective}
%
\author{D.M. Alexander\inst{1}
\and H. Aussel\inst{2}}
\authorrunning{Alexander and Aussel}
%
%
\institute{International School for Advanced Studies, SISSA,
Trieste, Italy
\and Osservatorio Astronomico di Padova, Padova, Italy}

\maketitle              

\begin{abstract}

We present new techniques to produce IRAS 12 $\mu$m samples of galaxies and stars. We
show that previous IRAS 12 $\mu$m samples are incompatible for detailed comparison
with ISO surveys and review their problems. We provide a stellar infrared diagnostic
diagram to distinguish galaxies from stars without using longer wavelength IRAS colour
criteria and produce complete 12 $\mu$m samples of galaxies and stars. This new
technique allows us to estimate the contribution of non-dusty galaxies to the IRAS 12
$\mu$m counts and produce a true local mid-infrared extragalactic sample compatible
with ISO surveys. We present our initial analysis and results.

\end{abstract}

\section{The importance of the local infrared picture} 

The recent ISO mission has produced a number of deep mid-infrared extragalactic
surveys [1,2,3,4] many of which are presented elsewhere in these proceedings.
In order to accurately evaluate the apparent source evolution found in these surveys
it is essential to have a stable and exact local infrared picture that is compatible
with ISO surveys.

\section{Previous work and problems}

There have been a number of previous IRAS 12 $\mu$m extragalactic samples produced
from either the Point Source Catalog, hereafter PSC, or the Faint Source Catalog,
hereafter FSC. The FSC was constructed by co-adding the individual PSC scans and is
consequently deeper at 12 $\mu$m by approximately one mag; the FSC is considered
complete to $f_{12}$$>$0.2 Jy. Due to the greater depth of the FSC only those samples
constructed from it will be considered here [5,6 hereafter RMS and FSXH]. In essence
none of these samples are truly compatible with the deeper ISO surveys because they
apply longer wavelength IRAS colour selection criteria and do not objectively classify
galaxies.  Some of these samples additionally suffer from inaccurate source flux
estimation, no correction for the overdensity due to large scale structure and
inaccurate K-correction. Due to the lack of space here these latter two points are not
considered although we refer the interested reader to [6,14] for excellent coverage of
these problems.

\subsection{The colour selection problem}

Selecting objects at 12 $\mu$m without colour selection will produce an abundance of
stars over galaxies due to the Jeans tail of stellar emission. Without exception every
extragalactic 12 $\mu$m sample to date has had (the majority of) stars removed by
applying longer wavelength IRAS colour criteria. This technique is clearly
incompatible with ISO surveys where no colour criterium is applied and will cause a
bias towards dusty galaxies. This also enforces that every galaxy must have a longer
wavelength flux, producing incompleteness even within the selection boundaries. For
example, in RMS the primary selection is $f_{12}$$>$0.22 Jy but every galaxy must also
have $f_{60}$$>$0.5$f_{12}$ or $f_{100}$$>$$f_{12}$. However due to the completeness
of FSC ($f_{60}$$>$0.2 Jy and $f_{100}$$>$0.6 Jy) this sample cannot be complete for
$f_{12}$$<$0.4 Jy or $f_{12}$$<$0.6Jy respectively.

\subsection{The classification problem}

To produce accurate extragalactic luminosity functions and understand the galaxy
contributions to fainter source counts it is necessary to classify galaxies in an
objective way, the most common technique is with optical line ratios [7,8]. To date
the only classified 12 $\mu$m sample is RMS although their classification was taken
from various catalogues which differ in the definition of extragalactic type and
completeness. As a comparison to this classification we have obtained line ratios from
the literature for 349 of the 483 RMS galaxies with $\delta$$>$0 degrees. This gives
completenesses of 72\%, 78\% and 93\% for objects $f_{12}$$>$0.22, 0.3 and 0.5 Jy
respectively. Due to the spectroscopic incompleteness at lower fluxes and the colour
selection incompleteness we only consider those of $f_{12}$$>$0.5 Jy here, see table
1; our classification follows that of [7,9].

\begin{table}
\centering
\caption{Extragalactic classification}
\renewcommand{\arraystretch}{1.4}
\setlength\tabcolsep{5pt}
\begin{tabular}{llllll}
\hline\noalign{\smallskip}
 & AGN & LINER & HII \\
\noalign{\smallskip}
\hline
\noalign{\smallskip}
RMS & 13\% & 15\% & - \\
AA & 16\% & 24\% & 60\% \\
\hline
\end{tabular}
\label{Tab1a}
\end{table}

All galaxies are found to show H$\alpha$ emission although for some galaxies
W$_{\lambda}$(H$\alpha$)$<$1 angstrom and they would appear as absorption line objects
in lower resolution/signal to noise spectra. Of the HII galaxies, 50\% show evidence
for significant star formation (W$_{\lambda}$(H$\alpha$)$>$10 angstroms). RMS
classified an object as an AGN if it is present in an AGN catalogue. We find a good
agreement in classification for this object class, the principal reason for the
construction of the RMS sample. LINERs were classified in the RMS sample if they were
present in an AGN catalgoue and consequently this sample is thought to be incomplete. 
We confirm this here. RMS did not classify HII galaxies although they considered those
galaxies not classified as a LINER or AGN, but with high infrared luminosities, to be
starburst galaxies and all other objects to be normal galaxies. 

\subsection{The source flux problem}

The FSC detection algorithm has been optimised for unresolved sources therefore fluxes
for extended sources need to be calculated from the coaddition of scans (the ADDSCAN
technique nowadays accessed via XSCANPI) [10]. However, if a source is unresolved, the
flux calculation using this technique leads erroneously to a larger flux than the FSC
flux [10]. Whilst FSXH carefully calculate the fluxes of extended and unresolved
sources seperately, RMS treat all sources as extended and consequently overestimate
the fluxes of unresolved galaxies; due to the large beamsize of IRAS a large number of
galaxies can be overestimated. The magnitude of this effect can be estimated from
ISOCAM observations. Unfortunately only a few observations are available for the lw10
filter (the closest to the IRAS 12 $\mu$m band) therefore in order to predict the IRAS
12 $\mu$m fluxes we have used observations in the lw2 (6.7 $\mu$m) and lw3 (14.3
$\mu$m) bands and a spectral decomposition technique similar to that described in
[11]. In this simple model the mid-infrared emission is produced by two components:
HII regions (using M17 [12]) and photo-dissociation regions (PDR)  (using NGC7023
[13]). The ratio of HII to PDR is calculated from the ratio of lw2 to lw3 fluxes, a
synthetic spectra is produced and the IRAS 12 $\mu$m flux is calculated. This
technique will be somewhat imprecise due to the uncertainty of the CAM photometry and
the ability of the model to reproduce the galactic spectrum.  Overall we estimate an
uncertainty of $\sim$20\%, roughly equal to the worst error in the FSC photometry. In
figure 1 we plot our predicted fluxes against the FSC and RMS fluxes divided by the
predicted flux (i.e.  the relative errors in flux).

\begin{figure}
\centering
\includegraphics[width=.6\textwidth]{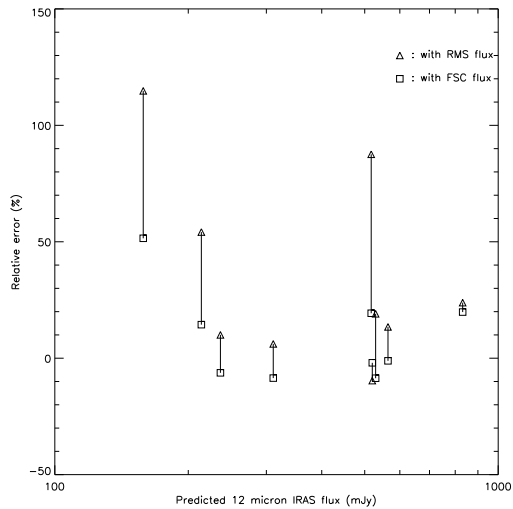}
\caption[]{12 $\mu$m flux comparison}
\label{eps1}
\end{figure}

Only those galaxies with z$>$2,500 kms$^{-1}$ have been plotted as they should all be
unresolved in the IRAS beam. In general a good agreement between the predicted flux
and FSC flux is found leading us to believe that the true fluxes for these objects are
close to the FSC flux. Consequently RMS may be overpredicting the source flux for
$\sim$50\% of their objects. This will have the effect of shifting galaxies from faint
flux/luminosity bins to higher bins causing unpredictable effects in the source count
and luminosity function determinations.

\section{A new IRAS 12 $\mu$m sample}

Our new IRAS 12 $\mu$m sample aims to address these problems and create a local
infrared sample that is compatible with ISO surveys. The sample definition is close to
that of RMS: sources are taken from the FSC for those objects with $|$b$|$$>$25
degrees, $f_{12}$$>$0.2 Jy and a moderate or good quality detection (S/N$>$3). This
selection results in 31002 objects, by comparison the RMS colour selection bias
produces $\sim$1100 objects. 

Our sample provides an interesting complement to the PSC-z extragalactic survey [15]
which selects those objects from the PSC with $f_{60}$$>$0.6 Jy. As with our sample
they do not apply colour selection criteria, stars are identified on Schmidt plates.
In terms of the extragalactic objects we would expect many galaxies in common although
our sample should have a higher fraction of quasars and early type galaxies.

\subsection{The stellar infrared diagnostic correlation}

By not applying colour selection to distinguish galaxies from stars we have had to
devise an alternative technique. The key assumption in our technique is that stars
have, in the majority, predictable properties and therefore with the large and
complete optical stellar databases currently available (in particular the Guide Star
Catalogue [16], hereafter GSC) it is possible to find the majority of stars in our
sample through positional cross correlation. An important factor here is in
determining the completeness of an optical stellar catalgoue in an infrared sample,
requiring an understanding of the general optical and infrared characteristics of
stars.
 
To determine these characteristics for our stars we have used SIMBAD, which provides
stellar classifications, and produced a large sample ($\sim$9000 objects) of
classified stars. When cross correlationing to optical positions we consider a star
correlated if its optical position falls within 5$\sigma$ of the IRAS position (the
mean major and minor error ellipse axes are 16.9" and 2.0" respectively). The
properties of our classified stellar sample are shown in figure 2. We have only
plotted those stars for which there are at least 10 objects in a classified class and
only a subset of these classes are shown here for clarity. A clear correlation between
stellar type and flux ratio is found. The predicted black body colours for the
different stellar types follows a straight line passing close to the A1V to K5III
points. The interesting deviation observed for stars beyond type MOIII is {\it
possibly} due to an increasing amount of stellar absorption in the V band.

\begin{figure}
\centering
\includegraphics[width=.6\textwidth]{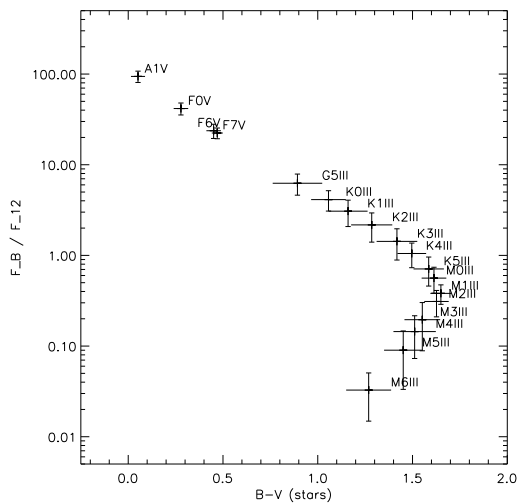}
\caption[]{Stellar infrared diagnostic diagram}
\label{eps1}
\end{figure}

These optical-infrared flux correlations provide an essential tool, the ability to
predict the IRAS 12 $\mu$m flux of a star for a given spectral type or B-V colour. In
the cases where we find an IRAS source associated with both a galaxy and a star we
will be able to predict the flux from the star and therefore estimate the flux from
the galaxy.

\subsection{Cross correlating stars}

The GSC is considered complete for 6$<$V$<$15 mags over the whole sky. Based on our
stellar infrared diagnostic correlation this corresponds to a depth of $f_{12}$$>$0.2
Jy and therefore any stars earlier than a type M6III in our sample should be in the
GSC. However, to accurately cross correlate the positions between the GSC and our
sample requires accounting for proper motion. The Schmidt plates for the GSC were
taken in 1975 and 1982, by comparison the mean IRAS observation epoch is 1983.5. By
analysing a sub-sample of Hipparcos stars ($\sim$4000 objects, selected by sky area)
we find a mean proper motion of 0.03"yr$^{-1}$, with a one $\sigma$ maximum of
0.13"yr$^{-1}$ (the maximum in the whole catalogue is 6"yr$^{-1}$). Taking into
account the different observation epochs and the mean IRAS error ellipse, virtually
all our stars should fall within 5$\sigma$ of their GSC position. 

We initially searched for associations by selecting all objects within 2' of the IRAS
position (2' corresponds to $\sim$5$\sigma$ of the mean positional uncertainty in the
major axis direction) although we only consider a star cross-correlated if it falls
within 5$\sigma$ of an IRAS source. From this cross correlation we find that over that
29000 of our IRAS sources are stars. Of these a small ($\sim$0.5\%), but significant,
fraction show a considerable infrared excess ($f_{25}$$>$$f_{12}$) and warrant further
study.

\subsection{Cross correlating galaxies}

Our extragalactic cross correlation is performed in a similar manner although the
problems associated with correlating galaxies are somewhat different. Proper motion is
not a problem although the extended size of galaxies is as the peak mid-infrared
position can vary from the peak optical position and therefore some positional
uncertainty must be included when trying to correlate an infrared galaxy to an optical
source. Although a 5$\sigma$ positional uncertainty can correspond to 2' if the galaxy
lies along the major axis of the IRAS error ellipse it can also correspond to just 10"
if the galaxy lies perpendicular to this direction. Therefore we only consider a
galaxy cross correlated if it falls within 10$\sigma$ of an IRAS source. Due to the
often unknown incompleteness of extragalactic catalogues it is not possible to
accurately determine the completeness of our sample in extragalactic catalogues and
therefore we have simply used the largest (and most appropriate) databases. In the
cross correlation presented here we have used the QIGC sample [17], which has just
become available in electronic form, NED, SIMBAD and the FSC 25 $\mu$m sample [18].

Using these databases we find that over 700 of our sources, not confirmed as stars,
fall within our 10$\sigma$ threshold. From these confirmed galaxies we find $\sim$50
are elliptical or S0 systems (not all are active galaxies) and $\sim$150 have
$f_{12}$$>$$f_{25}$ and are most probably PDR dominated galaxies. A number of galaxies
have $f_{12}$$>$2$f_{60}$ and would therefore not be picked up by RMS. One of these
galaxies is NGC 3115, a nearby bulge dominated galaxy of Hubble type S0. This galaxy
is only detected at 12 $\mu$m, the 25, 60 and 100 $\mu$m fluxes are upper limits,
although with a V band mag of $\sim$8.9 it is bright optical galaxy. In terms of it's
B-V (1.0) and B/12 (2.5) colours it corresponds to a K0III star. As a comparison we
have plotted a sample of our {\it normal} infrared galaxies, see figure 3.

\begin{figure}
\centering
\includegraphics[width=.6\textwidth]{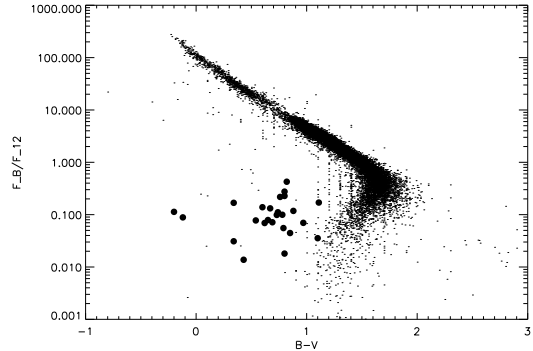}
\caption[]{Stellar and galactic colour plot. All confirmed stars are plotted as dots
and a selection of normal infrared galaxies are plotted as filled circles. NGC 3115 is
not plotted here but would correspond to the position of a K0III star, see figure 2}
\label{eps1} 
\end{figure}

\subsection{Further improvements}

Although our initial cross correlation has been successful we have $\sim$500 objects
for which we do not have an optical identification. These objects do not have IRAS
Cirrus/Confused flags or low 12 $\mu$m fluxes although approximately 75\% have upper
limit 60 $\mu$m IRAS fluxes. Visual inspection of a number of these objects with the
Digital Sky Survey shows them to be nearby bright stars, suggesting incompleteness in
the GSC at bright fluxes. However there is also probably a substantial population of
stars not yet accounted for: those with V$>$15 mags (e.g.  M7III and M8III stars),
dark molecular clouds and planetary nebulae. We are currently compiling a list of
additional stellar objects to cross correlate to our sample to allow us to produce a
definitive list of unidentified extragalactic sources.

\section{Further work}

Our primary aim is to construct complete 12 $\mu$m FSC selected samples of galaxies
and stars. From this we will create accurate extragalactic source counts and
classified luminosity functions with optical slit spectroscopy. As many of our
extragalactic objects will be extended we also intend to obtain integrated spectra to
provide a classification which is compatible with the distant objects found in ISO
surveys where the observed slit spectra will be produced by the majority of the
galaxy. With our stellar sample we wish to create a complete list of high galactic
latitude stars and sources to help constrain galactic models and provide further
diagnostics in distinguishing stars from galaxies in faint surveys (e.g. the Hubble
Deep Field [3]). Both our galactic and stellar samples show objects that deviate from
the norm (i.e. galaxies with stellar colours and stars with galactic colours) and
warrant further study in their own right.
\\~\\ 
{\bf Acknowledgements} We thank I.Matute for assistance in producing the initial
catalogue and M.Moshir for fruitful discussions. We are grateful to O.Laurent and
H.Roussel for providing the ISOCAM fluxes for some of the RMS sample sources. We
acknowledge and thank the EC TMR extragalactic networks for postdoctoral grant
support. This research has made use of the NASA/IPAC Extragalactic Database (NED)
which is operated by the Jet Propoulsion Laboratory, California Institute of
Technology, under contact with NASA.

\clearpage
\addcontentsline{toc}{section}{Index}
\flushbottom
\printindex

\end{document}